\documentclass[aps,prc,twocolumn,superscriptaddress,showpacs,amsmath,amssymb,floatfix]{revtex4}
\usepackage{graphicx}
\usepackage{dcolumn}
\usepackage{bm}
\usepackage{epsfig,amsmath,subfigure,xspace}
\usepackage{color}

\def\set{\sigma_{et}}
\def\se{\sigma_{e}}
\def\st{\sigma_{t}}
\def\so{\sigma_{0}}
\def\deg{^\circ}

\def\Cerenkov{{\v C}erenkov }
\def\gevm{{\rm GeV}}
\def\gevmsq{{\rm GeV}^2}
\def\gev{{\rm GeV}}
\def\gevpsq{{\rm GeV}^2}
\def\carb{${}^{12}\rm{C}$\xspace}
\def\nitr{${}^{15}\rm{N}$\xspace}
\def\nh{${}^{15}\rm{NH}_3$\xspace}
\def\nd{${}^{15}\rm{ND}_3$\xspace}

\def\hel{${}^{4}\rm{He}$\xspace}
\def\delt{$\Delta(1232)$\xspace}
\def\npp{N_{\uparrow\uparrow}}
\def\npm{N_{\uparrow\downarrow}}
\def\nmp{N_{\downarrow\uparrow}}
\def\nmm{N_{\downarrow\downarrow}}
\def\nppo{\overline{N_{\uparrow\uparrow}}}
\def\npmo{\overline{N_{\uparrow\downarrow}}}
\def\nmpo{\overline{N_{\downarrow\uparrow}}}
\def\nmmo{\overline{N_{\downarrow\downarrow}}}

\def\rea{{\rm Re}}
\def\reopmop{\rea\{E_{1+}^*M_{1+}\}}

\def\qA{$Q^2=0.252\ \gevpsq$}
\def\qB{$Q^2=0.385\ \gevpsq$}

\begin{document}


\title{First measurement of target and double spin asymmetries for $\vec{e}\vec{p} \rightarrow e p \pi^0$ in the nucleon resonance region above the $\Delta(1232)$}




\newcommand*{\ANL}{Argonne National Laboratory, Argonne, Illinois 60439}
\affiliation{\ANL}
\newcommand*{\ASU}{Arizona State University, Tempe, Arizona 85287-1504}
\affiliation{\ASU}
\newcommand*{\UCLA}{University of California at Los Angeles, Los Angeles, California  90095-1547}
\affiliation{\UCLA}
\newcommand*{\CSU}{California State University, Dominguez Hills, Carson, CA 90747}
\affiliation{\CSU}
\newcommand*{\CMU}{Carnegie Mellon University, Pittsburgh, Pennsylvania 15213}
\affiliation{\CMU}
\newcommand*{\CUA}{Catholic University of America, Washington, D.C. 20064}
\affiliation{\CUA}
\newcommand*{\SACLAY}{CEA-Saclay, Service de Physique Nucl\'eaire, 91191 Gif-sur-Yvette, France}
\affiliation{\SACLAY}
\newcommand*{\CNU}{Christopher Newport University, Newport News, Virginia 23606}
\affiliation{\CNU}
\newcommand*{\UCONN}{University of Connecticut, Storrs, Connecticut 06269}
\affiliation{\UCONN}
\newcommand*{\ECOSSEE}{Edinburgh University, Edinburgh EH9 3JZ, United Kingdom}
\affiliation{\ECOSSEE}
\newcommand*{\FU}{Fairfield University, Fairfield, CT 06824}
\affiliation{\FU}
\newcommand*{\SANTA}{
Universidad T\'ecnica Federico Santa Mar\'{\i}a,  Casilla 110-V, Valpara\'\i so, Chile}
\affiliation{\SANTA}
\newcommand*{\FIU}{Florida International University, Miami, Florida 33199}
\affiliation{\FIU}
\newcommand*{\FSU}{Florida State University, Tallahassee, Florida 32306}
\affiliation{\FSU}
\newcommand*{\GWU}{The George Washington University, Washington, DC 20052}
\affiliation{\GWU}
\newcommand*{\ECOSSEG}{University of Glasgow, Glasgow G12 8QQ, United Kingdom}
\affiliation{\ECOSSEG}
\newcommand*{\ISU}{Idaho State University, Pocatello, Idaho 83209}
\affiliation{\ISU}
\newcommand*{\INFNFR}{INFN, Laboratori Nazionali di Frascati, 00044 Frascati, Italy}
\affiliation{\INFNFR}
\newcommand*{\INFNGE}{INFN, Sezione di Genova, 16146 Genova, Italy}
\affiliation{\INFNGE}
\newcommand*{\ORSAY}{Institut de Physique Nucleaire ORSAY, Orsay, France}
\affiliation{\ORSAY}
\newcommand*{\ITEP}{Institute of Theoretical and Experimental Physics, Moscow, 117259, Russia}
\affiliation{\ITEP}
\newcommand*{\JMU}{James Madison University, Harrisonburg, Virginia 22807}
\affiliation{\JMU}
\newcommand*{\KYUNGPOOK}{Kyungpook National University, Daegu 702-701, Republic of Korea}
\affiliation{\KYUNGPOOK}
\newcommand*{\UMASS}{University of Massachusetts, Amherst, Massachusetts  01003}
\affiliation{\UMASS}
\newcommand*{\MOSCOW}{Moscow State University, General Nuclear Physics Institute, 119899 Moscow, Russia}
\affiliation{\MOSCOW}
\newcommand*{\UNH}{University of New Hampshire, Durham, New Hampshire 03824-3568}
\affiliation{\UNH}
\newcommand*{\NSU}{Norfolk State University, Norfolk, Virginia 23504}
\affiliation{\NSU}
\newcommand*{\OHIOU}{Ohio University, Athens, Ohio  45701}
\affiliation{\OHIOU}
\newcommand*{\ODU}{Old Dominion University, Norfolk, Virginia 23529}
\affiliation{\ODU}
\newcommand*{\RPI}{Rensselaer Polytechnic Institute, Troy, New York 12180-3590}
\affiliation{\RPI}
\newcommand*{\RICE}{Rice University, Houston, Texas 77005-1892}
\affiliation{\RICE}
\newcommand*{\URICH}{University of Richmond, Richmond, Virginia 23173}
\affiliation{\URICH}
\newcommand*{\SCAROLINA}{University of South Carolina, Columbia, South Carolina 29208}
\affiliation{\SCAROLINA}
\newcommand*{\JLAB}{Thomas Jefferson National Accelerator Facility, Newport News, Virginia 23606}
\affiliation{\JLAB}
\newcommand*{\UNIONC}{Union College, Schenectady, NY 12308}
\affiliation{\UNIONC}
\newcommand*{\VT}{Virginia Polytechnic Institute and State University, Blacksburg, Virginia   24061-0435}
\affiliation{\VT}
\newcommand*{\VIRGINIA}{University of Virginia, Charlottesville, Virginia 22901}
\affiliation{\VIRGINIA}
\newcommand*{\WM}{College of William and Mary, Williamsburg, Virginia 23187-8795}
\affiliation{\WM}
\newcommand*{\YEREVAN}{Yerevan Physics Institute, 375036 Yerevan, Armenia}
\affiliation{\YEREVAN}
\newcommand*{\NOWJLAB}{Thomas Jefferson National Accelerator Facility, Newport News, Virginia 23606}
\newcommand*{\NOWUNH}{University of New Hampshire, Durham, New Hampshire 03824-3568}
\newcommand*{\NOWSACLAY}{CEA-Saclay, Service de Physique Nucl\'eaire, 91191 Gif-sur-Yvette, France}
\newcommand*{\NOWGWU}{The George Washington University, Washington, DC 20052}
\newcommand*{\NOWUMASS}{University of Massachusetts, Amherst, Massachusetts  01003}
\newcommand*{\NOWMIT}{Massachusetts Institute of Technology, Cambridge, Massachusetts  02139-4307}
\newcommand*{\NOWECOSSEE}{Edinburgh University, Edinburgh EH9 3JZ, United Kingdom}
\newcommand*{\NOWSCAROLINA}{University of South Carolina, Columbia, South Carolina 29208}
\newcommand*{\NOWOHIOU}{Ohio University, Athens, Ohio  45701}
\author {A.S.~Biselli} 
\email{biselli@jlab.org}
\thanks{Corresponding author.}
\affiliation{\FU}
\author {V.D.~Burkert} 
\affiliation{\JLAB}
\author {M.J.~Amaryan} 
\affiliation{\ODU}
\author{G.~Asryan} 
\affiliation{\YEREVAN}
\author {H.~Avakian} 
\affiliation{\JLAB}
\author {H.~Bagdasaryan} 
\affiliation{\ODU}
\author {N.~Baillie} 
\affiliation{\WM}
\author {J.P.~Ball} 
\affiliation{\ASU}
\author {N.A.~Baltzell} 
\affiliation{\SCAROLINA}
\author {M.~Battaglieri} 
\affiliation{\INFNGE}
% ...

%
%
\date{\today}

\begin{abstract}
The exclusive channel $\vec{p}(\vec{e},e'p)\pi^0$  
was studied in the first and second nucleon resonance regions in the $Q^2$ range from $0.187$ to $0.770\ \gevpsq$ at Jefferson Lab using the CEBAF Large Acceptance Spectrometer (CLAS). 
Longitudinal target and beam-target asymmetries were extracted over a large range of  center-of-mass angles of the $\pi^0$
and compared to  the unitary isobar model MAID, the dynamic model by Sato and Lee, and the dynamic model DMT.
A strong sensitivity to individual models was observed, in particular for the target asymmetry and in the higher invariant mass region. 
This data set, once included in the global fits of the above models,
is expected to place strong constraints on the electrocoupling
amplitudes $A_{1/2}$ and $S_{1/2}$ for the Roper resonance $N(1400)P_{11}$, and the 
$N(1535)S_{11}$ and $N(1520)D_{13}$ states. 
\end{abstract}


\maketitle

\section{Introduction}

Studying the structure of the nucleon and its excitations is essential in the development of an understanding of the strong interaction at the confinement scale \cite{isgur}. On a fundamental level there exists only a limited understanding of the relationship between Quantum-Chromo-Dynamics (QCD), the field theory of the strong interaction, and models that use effective degrees of freedom, such as the large variety of constituent quark models or alternative hadron models \cite{dmt,drechsel99,lee01,IGAzn,Aznauryan2005_1}. Experimentally, we still lack sufficiently complete data sets that can be used in phenomenological analyses to unravel the internal structure of the nucleon's excited states by determining their electromagnetic transition amplitudes for different distance scales. Precise cross section and polarization data for $\pi^0$ production off protons, which are needed to study the transition from the nucleon ground state to the $\Delta(1232)$, have become available in recent years. For a recent review see Ref.~\cite{bur-lee04}. These data have been used to extract the electromagnetic transition multipoles for this state over  a wide range of  photon virtuality $Q^2$ \cite{frolov99,joo02,joo03,joo04,sparveris06,biselli03,kelly05,ungaro06,stave06}. These accurate data have been confronted with recent results from quenched Lattice QCD calculations\cite{alexandrou04,alexandrou05}. 

The $\Delta(1232)P_{33}$ excitation corresponds to the transition between isospin 1/2 and isospin 3/2 ground states of the non-strange baryons and is characterized by a spin flip in the ground state. To study the radial and orbital excitations of the nucleon, we need to measure the transition form factors for higher mass states. In the mass region above the $\Delta(1232)P_{33}$, there is a cluster of resonances around 1.5~GeV that consists of 3 isospin 1/2 nucleon excitations. The $N(1440)P_{11}$, often called the Roper resonance, is thought of as the first radial excitation of the nucleon. The two other states, the $N(1520)D_{13}$ and $N(1535)S_{11}$, are orbital excitations of the nucleon. By measuring the transition form factors of these states, we probe the radial and orbital wave function of the nucleon 
that describe
fundamental properties of the nucleon structure at the constituent quark and meson-nucleon scales. In the mass range around 1.7~GeV, there is a cluster of nine resonances that consists of a number of isospin 1/2 and isospin 3/2 states in various configurations of radial and orbital excitations of the 3-quark system.
 These resonances overlap with each other and contain higher partial waves. This mass range is also characterized by significant non-resonant contributions to the production cross section. In order to separate these states, and isolate them from non-resonant contributions, more experimental information is required than in the case of the $\Delta(1232)$. Measurements of different isospin channels, e.g. $p\pi^0$ and $n\pi^+$ final states, are essential. Measurements with polarized beam and polarized target, and measurements of the proton recoil polarization, are highly sensitive to interferences of resonant and non-resonant amplitudes contributing to pion production. These data  are indispensable for isolating individual resonances and determining their excitation strengths, within the framework of any model. 

The mass range above the $\Delta(1232)$ has recently been systematically mapped out in measurements of unpolarized cross sections of the electroproduction processes $ep\rightarrow en\pi^+$ \cite{egiyan06,park2007}, $ep\rightarrow ep\pi^0$ \cite{joo02}, $ep\rightarrow ep\eta$ \cite{armstrong99,thompson01,denizli07}, and $ep\rightarrow ep\pi^+\pi^-$ \cite{ripani03,strauch}, and measurements of beam spin asymmetries using polarized electron beams \cite{egiyan06,joo05}. Detailed measurements with polarized beams and polarized targets were carried out in inclusive electron scattering in the resonance region \cite{renee,yun,bosted}, and a limited number of data points were measured for the $\vec{e}\vec{p} \rightarrow en\pi^+$ reaction with polarized proton targets \cite{devita01}. However, no data are available for the processes ${e}\vec{p} \rightarrow ep\pi^0$ and $\vec{e}\vec{p} \rightarrow ep\pi^0$. The intent of the current research has been to fill this gap and to provide precise and detailed data from measurements for the $p\pi^0$ final state using both polarized electrons and a polarized proton target, and to systematically map the mass region from $W=1.1$ GeV to $W=1.7$ GeV and in the range  $0.187 < Q^2 < 0.770$ GeV$^2$.

\section{Formalism}
\begin{figure}[here,top]
\includegraphics[width=8.5cm]{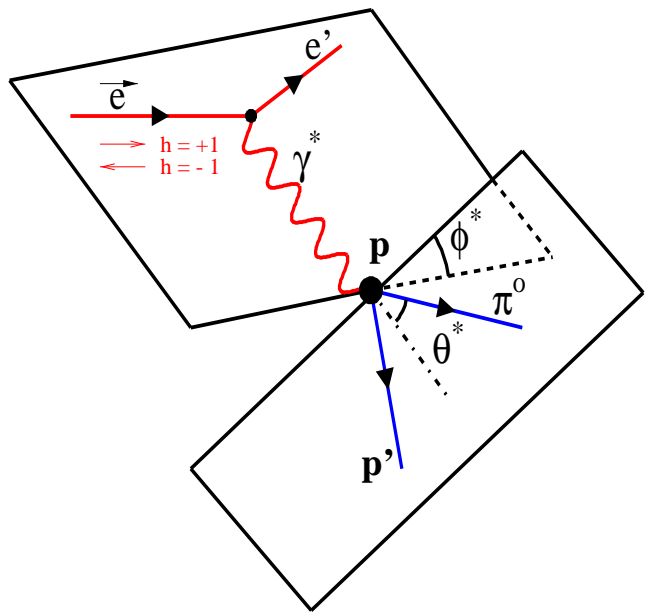}
\caption{(Color online) Schematic diagram of $\pi^0$-nucleon electro-production.
 $\vec{e}$ represents the incident polarized electron, $e'$ is the outgoing electron,
 $\gamma^{*}$ is the virtual photon, and $p$ and $p'$ are the nucleon in the initial
 and final states, respectively.}
\label{elecprod}
\end{figure}

The $\vec{e}\vec{p} \rightarrow ep\pi^0$ reaction is schematically shown in
Figure~\ref{elecprod}.
The incident polarized electron is given by the
 4-vector $p_e=(\vec{p}_e,E_i)$, the outgoing electron is emitted with
 angles $\phi_{e}$, $\theta_{e}$ and 4-vector
$p_e'=(\vec{p}_e\,',E_{f})$,  and
 the nucleon initial and final states are given by $p_p=(0,M)$ and
 $p_p'=(\vec{p}_p',E_p)$, respectively. The virtual photon $\gamma^*$ with 4-vector $q=(\vec{q},\omega)$ where $\vec{q}=\vec{p}_e-\vec{p}_e\,'$ and $\omega=E_i-E_f$, is characterized by the squared four momentum transfer:
\begin{equation}
Q^2=-q^2=-(\omega^2-{\mid\vec{q}\mid}^2)=4E_f E_i\sin^2\frac{\theta_e}{2}
\end{equation}
and the degree of transverse polarization:
\begin{equation}
\epsilon=\biggl(1+2\frac{{\mid\vec{q}\mid}^2}{ Q^2}\tan^2\frac{\theta_e}{2}\biggr)^{-1}.
\end{equation}
The invariant mass of the hadronic system is given by:
\begin{equation}
W=|p_e+p_p-p_e'|=\sqrt{M^2+2M\omega-Q^2}.
\end{equation}

In terms of these variables
 the differential cross section can be written as:
\begin{equation}
\frac {d^5\sigma}{d E_f d\Omega_e d\Omega^*}=\Gamma\frac{d\sigma}{d\Omega ^*},
\end{equation}
where
$d\Omega_e=\sin \theta_e d\theta_e d\phi_e $ is the electron solid angle and
$d\Omega^*=\sin \theta^* d\theta^* d\phi^*$ is the solid angle of the meson in the center of mass.
Here
\begin{equation}
\Gamma=\frac{\alpha}{2\pi^2}\frac{E_f}{E_i}\frac{k^{lab}_{\gamma}}{Q^2}\frac{1}{
1-\epsilon}
\end{equation}
is the virtual photon flux, where
\begin{equation}
k^{lab}_{\gamma}=\frac{W^2-M^2}{2M}
\end{equation}
denotes the `photon equivalent energy' necessary for a real photon
to excite a hadronic system with center-of-mass (c.m.) energy $W$,
and $\alpha$ is the fine structure constant.

The differential cross section for pion production by a virtual
photon $d\sigma / d\Omega^{*}$ can be written as a sum of four terms:
\begin{equation}
\label{forsigma2}
\frac{d\sigma}{d\Omega^{*}}= \frac{|\vec{k}|}{k_{\gamma}^{c.m.}}\biggl\{\frac{d\sigma_0}{d\Omega^{*}}+h\frac{d\sigma_e}{d\Omega^{*}}+P\frac{d\sigma_t}{d\Omega^{*}}+hP\frac{d\sigma_{et}}{d\Omega^{*}}\biggr\},
\end{equation}
where $\vec{k}$ is the momentum of the pion, $h$ is the electron helicity, and $P$ is the target proton polarization.
The first term $d\sigma_0 / d\Omega^{*}$ represents the unpolarized cross section,
while the remaining terms $d\sigma_e / d\Omega^{*}$, $d\sigma_t / d\Omega^{*}$, and $d\sigma_{et} / d\Omega^{*}$
arise when beam, target, or both beam and target are polarized, respectively. Here
\begin{equation}
k_{\gamma}^{c.m.}=\frac{M}{W}k_{\gamma}^{lab}
\end{equation}
is the `real photon equivalent energy' in the c.m. frame.
These cross sections can be written in terms of response functions $R$ using the formalism of reference~\cite{drechsel92} as:
\begin{widetext}
\begin{equation}
\label{forsigma}
\begin{split}
\frac{d\sigma_0}{d\Omega^{*}}
        =&R_{T}^{0} + \epsilon_{L} R_{L}^{0} +\sqrt{2\epsilon_{L}(1+\epsilon)} R_{TL}^{0} \cos\phi^{*} + \epsilon R_{TT}^{0} \cos 2\phi^{*}\\
\frac{d\sigma_e}{d\Omega^{*}}
        =&\sqrt{2\epsilon_{L}(1-\epsilon)} R_{TL'}^{0} \sin\phi^{*}\\
\frac{d\sigma_t}{d\Omega^{*}}
        =&\sin\theta_{\gamma}\cos\phi^*[\sqrt{2\epsilon_{L}(1+\epsilon)} R_{TL}^{x} \sin\phi^{*} + \epsilon R_{TT}^{x} \sin 2\phi^{*}]
        +\sin\theta_{\gamma}\sin\phi^*[R_{TL}^{y} + \epsilon_{L} R_{L}^{y} +  \\
        +& \sqrt{2\epsilon_{L}(1+\epsilon)} R_{TL}^{y} \cos\phi^{*}+ \epsilon R_{TT}^{y} \cos 2\phi^{*}] +\cos\theta_{\gamma}[\sqrt{2\epsilon_{L}(1+\epsilon)} R_{TL}^{z} \sin\phi^{*} + \epsilon R_{TT}^{z} \sin 2\phi^{*}]\\
\frac{d\sigma_{et}}{d\Omega^{*}}
        =&-\sin\theta_{\gamma}[\sqrt{2\epsilon_{L}(1-\epsilon)} R_{TL'}^{x} {\cos\phi^{*}}^{2} + \sqrt{1-\epsilon^{2}} R_{TT'}^{x}\cos\phi^*]
        +\sin\theta_{\gamma}\sqrt{2\epsilon_{L}(1-\epsilon)} R_{TL'}^{y} {\sin\phi^{*}}^{2}\\
        -&\cos\theta_{\gamma} [\sqrt{2\epsilon_{L}(1-\epsilon)} R_{TL'}^{z} \cos\phi^{*} + \sqrt{1-\epsilon^{2}} R_{TT'}^{z}],\\
\end{split}
\end{equation}
\end{widetext}
where
\begin{equation}
\epsilon_L=\frac{Q^2}{\omega^2}\epsilon
\end{equation}
is the frame-dependent longitudinal polarization of the virtual photon. The $\theta_{\gamma}$ is the angle between the directions of the target polarization and the  virtual photon.

The asymmetries are then defined as:
\begin{equation}
\begin{split}
A_e=&\frac{\se}{\so}\\
A_t=&\frac{\st}{\so}\\
A_{et}=-&\frac{\set}{\so},\\
\end{split}
\end{equation}
where  $\so \equiv d\sigma_0 / d\Omega^{*}$, $\se \equiv d\sigma_e / d\Omega^{*}$, $\st \equiv d\sigma_t / d\Omega^{*}$, and $\set \equiv d\sigma_{et} / d\Omega^{*}$.

If integrated over the azimuthal center-of-mass angle $\phi^*$, the double spin asymmetry $A_{et}$ can be parameterized as~\cite{Filippone}:
\begin{equation}
A_{et}=\sqrt{1-\epsilon^2}\cos \theta_{\gamma} \frac{A_1+\eta A_2}{1+\epsilon R},
\end{equation}
where $\eta=\tan \theta_{\gamma}\sqrt{2\epsilon/(1+\epsilon)}$ and $R$ is the longitudinal-transverse cross section ratio $\sigma_L/\sigma_T$ The structure function $A_1$ is the virtual photon helicity asymmetry,
\begin{equation}
A_1=\frac{|A_{1/2}|^2-|A_{3/2}|^2}{|A_{1/2}|^2+|A_{3/2}|^2},
\end{equation}
while $A_2$ is a longitudinal-transverse interference term.
\section{Reaction Models \label{sec:models}}

In recent years, several models have been developed that aim at accurately reproducing the experimental data. In section~\ref{sec:results} we compare some of our data with calculations based upon model descriptions such as the Dubna-Mainz-Taipei~(DMT) model~\cite{dmt}, a recent version of the MAID model~\cite{drechsel99}, and the Sato-Lee model~(SL)~\cite{lee01}. In addition, a unitary isobar model (UIM) was developed by the Yerevan-JLab group~\cite{IGAzn,Aznauryan2005_1} that contains many features of MAID, but incorporates different energy-dependences of the background amplitudes at higher energies. 
This approach allows the fitting of experimental cross sections and polarization asymmetries and the extraction of resonance
transition amplitudes from a large body of data. We briefly summarize the main features of these models. They are discussed in more detail in Ref.~\cite{bur-lee04}.

\begin{figure}[htb]
\begin{center}
   \includegraphics[angle=0,width=8.5cm]{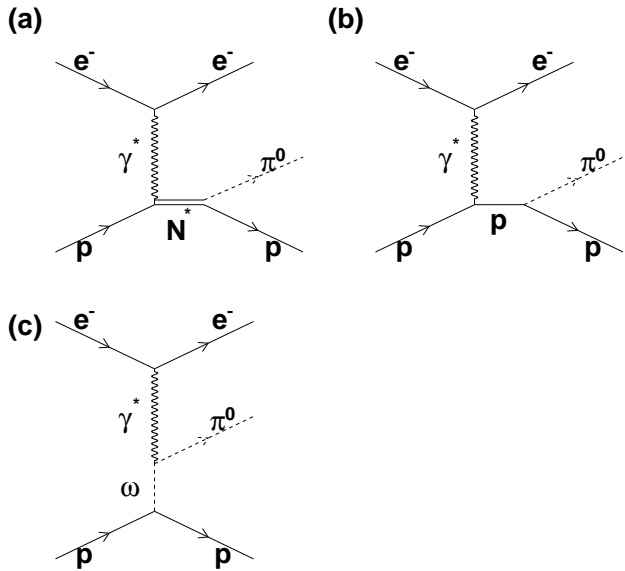}
       \caption{
         Tree-level diagrams of single $p\pi^0$ electroproduction for (a) $s-$channel resonance production, (b) $s-$channel nucleon exchange and, (c) $t-$channel $\omega$ meson exchange.}
         \label{fig:Born_terms}
\end{center}
\end{figure}

MAID and related models are based on an isobar description of the single pion production process. They incorporate non-resonant amplitudes described by tree-level Born terms, and also include $\rho$ and $\omega$ $t$-channel processes that are relevant mostly in the region of higher resonances. Figure~\ref{fig:Born_terms} shows the diagrams contributing to the reaction $e p \to e p\pi^0$ at low and intermediate energies. The vertex functions for the virtual photon coupling to hadrons are parameterized according to their respective on-shell form factors for which there is prior experimental information. Resonances are parameterized by a phenomenological description using a relativistic Breit-Wigner form with an energy-dependent width. The total amplitude for single pion production is unitarized in a K-matrix formulation. Only single channels are included, and multi-channel effects such as $\gamma N \to (\rho N, \pi\Delta) \to \pi N$, which could be important in the second and third resonance regions, are neglected. From an experimental viewpoint, the attractive feature of these descriptions is flexibility that allows adjusting parameters such as electromagnetic transition form factors and hadronic couplings as new experimental information becomes available. However, all of these descriptions lack significant predictive power, and a comparison with new data will tell us more about how well electromagnetic and hadronic couplings have been parameterized, rather than about the intrinsic structure of the nucleon.       

Dynamical models, such as the SL and DMT models start from a consistent Hamiltonian formulation. In these models, the non-resonant interaction modifies the resonant amplitude. The SL model provides the most consistent description of the interaction but is currently limited to the region of the $\Delta(1232)$ resonance, while in the DMT model the resonance amplitudes are parameterized according to a specific Breit-Wigner form that simplifies the inclusion of higher resonances. The $s$-channel resonance parameterization in the DMT model is similar to what is used in the isobar descriptions such as the MAID and UIM approaches. 
Once the transition form factors have been extracted from the data, their interpretation in terms of the intrinsic structure of the nucleon must then involve comparisons with nucleon structure models, such as the many versions of the Constituent Quark Model's (CQM), and Lattice QCD calculations. 

\section{Experimental setup}
\label{sec:expsetup}
Measurements were performed in 2000-2001 using the CEBAF Large Acceptance Spectrometer (CLAS)~\cite{Mecking03} at the Thomas Jefferson National Accelerator Facility (Jefferson Lab), located in Newport News, Virginia. 
The experiment consisted of a longitudinally polarized electron beam with energies from 1.6 $\gev$ to 5.6 $\gev$ and currents ranging from 1 to 5 nA, scattering off longitudinally polarized protons and deuterons in ammonia targets, \nh and \nd. The beam was rastered in a spiral pattern of 1-1.2 cm diameter over the target to avoid destroying the target polarization due to beam heating. The beam had an average polarization of 70\%, and the polarization direction was reversed with a frequency of 30 Hz. 

The experiment used two frozen ammonia targets,  \nh and \nd, and  \carb, \hel, and \nitr targets for background studies. The targets were held in a 1 K liquid-helium cooling bath and a 5 T longitudinal magnetic field generated by a pair of superconducting Helmholtz coils, which blocked particles scattered at polar angles between 45$\deg$ and 70$\deg$. The protons and deuterons in the ammonia were polarized via Dynamic Nuclear Polarization and the polarization achieved was about 75\% for protons and 25\%  deuterons. 

The scattered particles were detected using the CLAS detector, described in great detail in Ref.~\cite{Mecking03}. The CLAS detector consists of 6 superconducting coils that produce a toroidal field, around the beam direction, which bends charged particles in the polar direction but leaves them nearly unaffected in the azimuthal direction. 
The polarity of the magnetic field was set to bend negative particles toward the beam line.
Each of 6 sectors, delimited  by the coils, are equipped with three drift chamber (DC) regions to determine the trajectories of the charged particles, a {\v C}erenkov counter (CC) for electron identification, scintillators counters (SC) to measure the time-of-flight (TOF), and an electromagnetic calorimeter (EC) used to identify electrons and neutral particles. The detector covers 80\% of the azimuthal angle. The DC and  SC subtend polar angles from 8$\deg$ to 142$\deg$, while the CC and EC cover up to 45$\deg$.
 
\section{Data reduction and analysis}
In this paper we report the measurement of target and double spin asymmetries for the $\vec{p}(\vec{e},e'p)\pi_0$ channel using the 1.6 $\gev$ beam on \nh target data set.

\label{sec:analysis}
\subsection{Channel identification}
\begin{figure}
 \includegraphics[width=8.5cm]{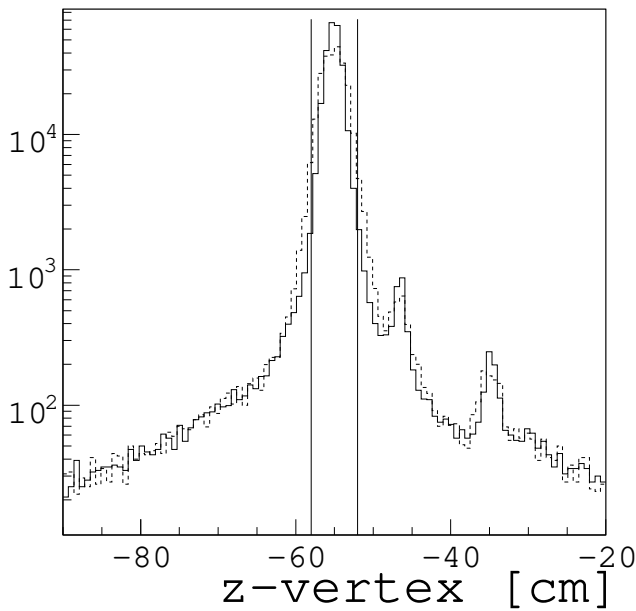}%
 \caption{\label{z-vertex}Number of electron events as a function of the vertex $z$-position. The dashed and the solid lines represent the distribution before and after the raster correction, respectively. The vertical lines indicate the cut applied to the data to remove the peaks coming from the scattering off the exit windows of the target.}
 \end{figure}

The exclusive final state $\vec{p}(\vec{e},e'p)\pi^0$ was selected by identifying an electron, a proton, and a missing pion.
Electrons are identified at the trigger level by requiring a coincidence signal from the CC and the EC, and  the offline track reconstruction reduces the electron candidates by geometrically matching signals from the DC, CC and EC.   Electron identification was further improved in the data analysis in order to reduce contamination from negative particles such as $\pi^-$ and other sources of background. The $z$-vertex position was used to remove events originating from scattering off the target windows. 
In order to improve the resolution of the vertex position, 
the data were corrected  to account for the effect on the vertex reconstruction due to the beam rastering. Figure~\ref{z-vertex} shows the vertex position before and after the correction, and the applied cut, $-58\ {\rm cm}<$ $z$-vertex $<-52\ {\rm cm}$, to select events originating inside the  \nh target. 

To separate electrons from negative pions, a combinations of cuts on the \Cerenkov and the EC signals were imposed.  
The energy released by electrons traversing the EC is proportional to their momentum, and about 30\% of this energy deposition is measured by the scintillator sheets in the EC (sampling fraction).  To reject negative pions, a cut  on the total EC visible energy as a function of the momentum was applied.  The energy deposited in the EC is also measured separately for the inner part and the outer part. A cut to reject events with a low energy deposition in the inner part of the calorimeter was applied. 
This allows further identification of electrons, which shower mostly in the inner part of the EC, while pions are minimum ionizing and lose most of their energy in the outer part.

\begin{figure}
 \includegraphics[width=8.5cm]{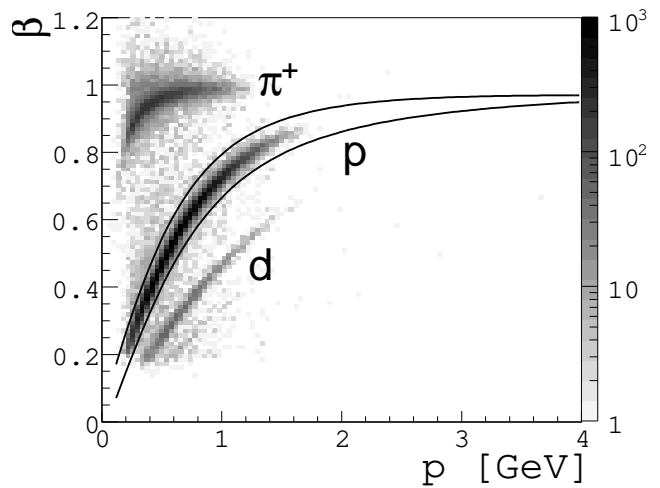}%
 \caption{\label{pospid} $\beta$ vs $p$ for positively charged particles. The lines show how protons are selected in the analysis.}
 \end{figure}

In CLAS, positive hadrons are identified using the momentum and the path length information from the DC and the timing information from the SC.
Figure~\ref{pospid} shows the cut applied to select protons, which appear well separated from positive pions at the energy of this experiment.
The missing pions were selected by a 2$\sigma$ cut on the missing mass $M_X^2=|p_e+p_p-p_{e'}-p_{p'}|^2$ spectrum around the mass of the pion. In addition, to reduce multi-pion background, 
events with detected charged particles other than the electron and the proton in the final state were rejected.
\subsection{Fiducial cuts and acceptance corrections}
The objective of this analysis is to extract asymmetries, for which the precise knowledge of the acceptance is not as crucial as for differential cross section measurements. Nevertheless, the acceptance can still affect the results. To calculate asymmetries, the ratio of counts integrated over ranges of $\phi^*$, $\theta^*$, $Q^2$, and $W$ is formed, and since the acceptance depends upon these variables, it does not necessarily cancel. Fiducial cuts were applied to the data in order to remove regions where the efficiency varies by more than one order of magnitude, and the acceptance was calculated analytically for the geometrical regions defined by the fiducial cuts. This procedure fully corrects the data for the acceptance except for an overall efficiency factor that is assumed to be constant within the fiducial cuts, and which will therefore cancel out when taking the ratios in the asymmetries.
\subsection{Experimental definition of the asymmetries}
For each helicity configuration, the number of events can be written as:
\begin{equation}
\begin{split}
\npp = D_1(\so+\so^N+P_e^a\se+P_e^a\se^N+P_t^a\st+P_e^aP_t^a\set)\\
\nmp = D_2(\so+\so^N-P_e^a\se-P_e^a\se^N+P_t^a\st-P_e^aP_t^a\set)\\
\npm = D_3(\so+\so^N+P_e^b\se+P_e^b\se^N-P_t^b\st-P_e^bP_t^b\set)\\
\nmm = D_4(\so+\so^N-P_e^b\se-P_e^b\se^N-P_t^b\st+P_e^bP_t^b\set),
\end{split}
\end{equation}
where the arrows in $\npp$, $\npm$, $\nmp$, and $\nmm$ indicate  the beam and the target polarizations, respectively.
Since the target was the same for all configurations, the parameters $D_1$ through $D_4$ differ only by the total  accumulated charge $Q$. $P_e^a$ and $P_t^a$ are the beam and target polarizations for runs with a positively polarized target, and $P_e^b$ and $P_t^b$ for negative target polarization runs. The beam  polarizations  for the groups ($\npp$,$\nmp$) and ($\npm$,$\nmm$) are assumed to be equal because the polarization of the beam was reversed every half second, and therefore each pair was taken at the same time. On the other hand, the target polarization was changed on a time scale of several days, and therefore the corresponding beam polarization could be different during these different periods. The terms $\so$, $\se$, $\st$, and $\set$ represent the cross sections for scattering of polarized electrons off a longitudinally polarized proton in the \nh target, while the terms $\so^N$ and $\se^N$ were added to take into account that there are additional contributions coming from the 7 unpolarized protons bound in the nitrogen nucleus.
The asymmetries may be written in terms of these quantities as:
\begin{eqnarray}
A_t&=\frac{\st}{\so}=\frac{1}{P_t^b}\frac{(\nppo+\nmpo)-(\npmo+\nmmo)}{(\nppo+\nmpo)+\alpha_t(\npmo+\nmmo)-C N ^{BG}}, \label{forasymdef}\\
A_{et}&=-\frac{\set}{\so}=\frac{1}{P_e^aP_t^b}\frac{-(\nppo-\nmpo)+\alpha_{e}(\npmo-\nmmo)}{(\nppo+\nmpo)+\alpha_t(\npmo+\nmmo)-C N^{BG}},
\label{forasymdef_bis}
\end{eqnarray}
where the counts were normalized to the same total charge. In these expressions,
\begin{equation}
\begin{split}
\nppo\equiv\frac{D_2}{D_1}\npp  \hspace{1cm}
\nmpo\equiv\nmp\\
\npmo\equiv\frac{D_2}{D_1}\npm   \hspace{1cm}
\nmmo\equiv\frac{D_2}{D_4}\nmm
\end{split}
\end{equation}
and
\begin{equation}
\label{forasymdef2}
\begin{split}
\alpha_{e}=\frac{P_e^a}{P_e^b} \hspace{1cm}
\alpha_{t}=\frac{P_t^a}{P_t^b}.\\
\end{split}
\end{equation}
The term $C \cdot N ^{BG}$ represents the properly normalized \carb and \hel counts used to simulate and remove the residual background contribution $2D_2(1+\alpha_t)\so^N$ in the denominator. The experimental determination of the normalization constant $C$ is discussed in the next section.
\subsection{Nuclear background subtraction}
\begin{figure}[top]
 \includegraphics[width=4.cm]{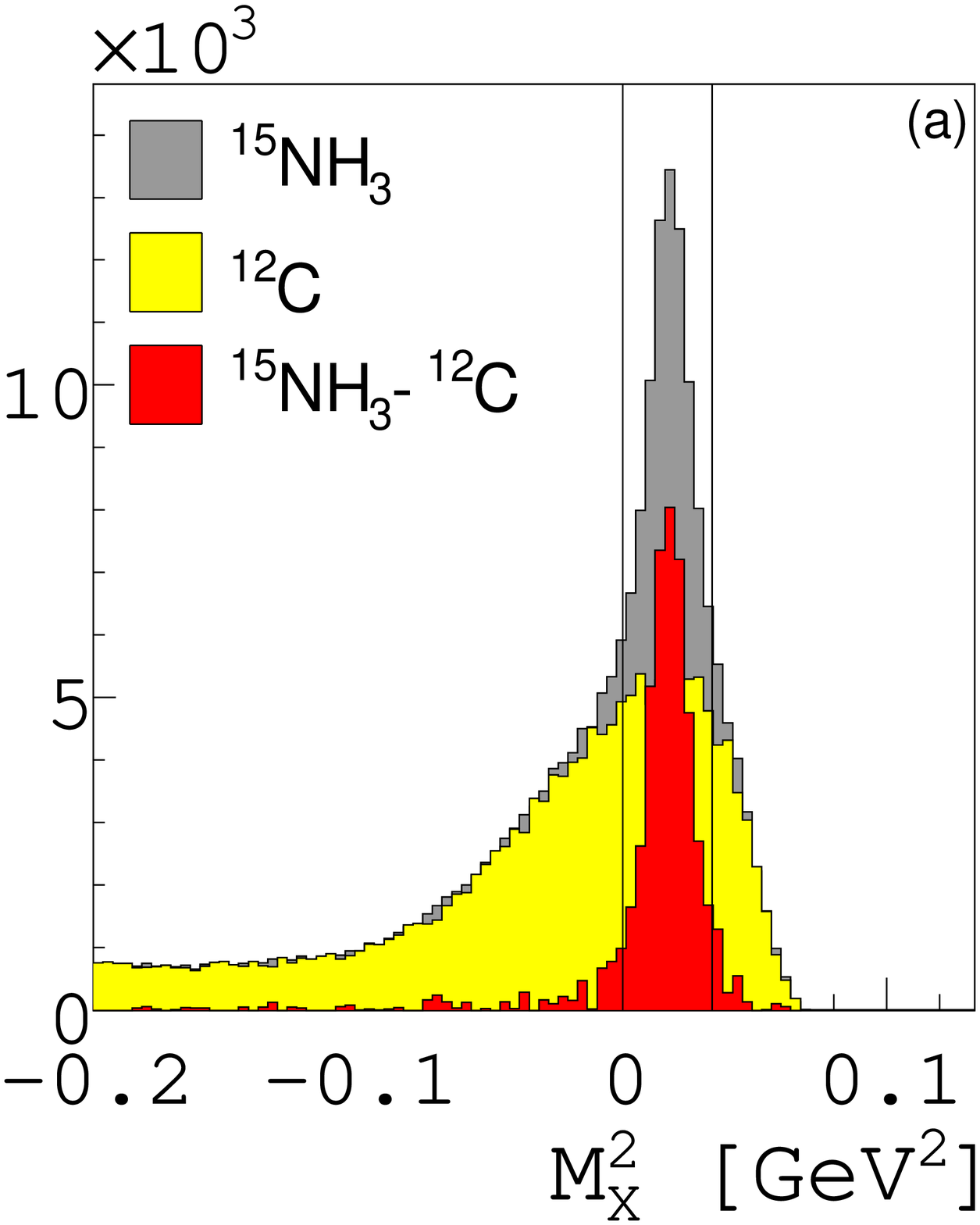}%
 \includegraphics[width=4.cm]{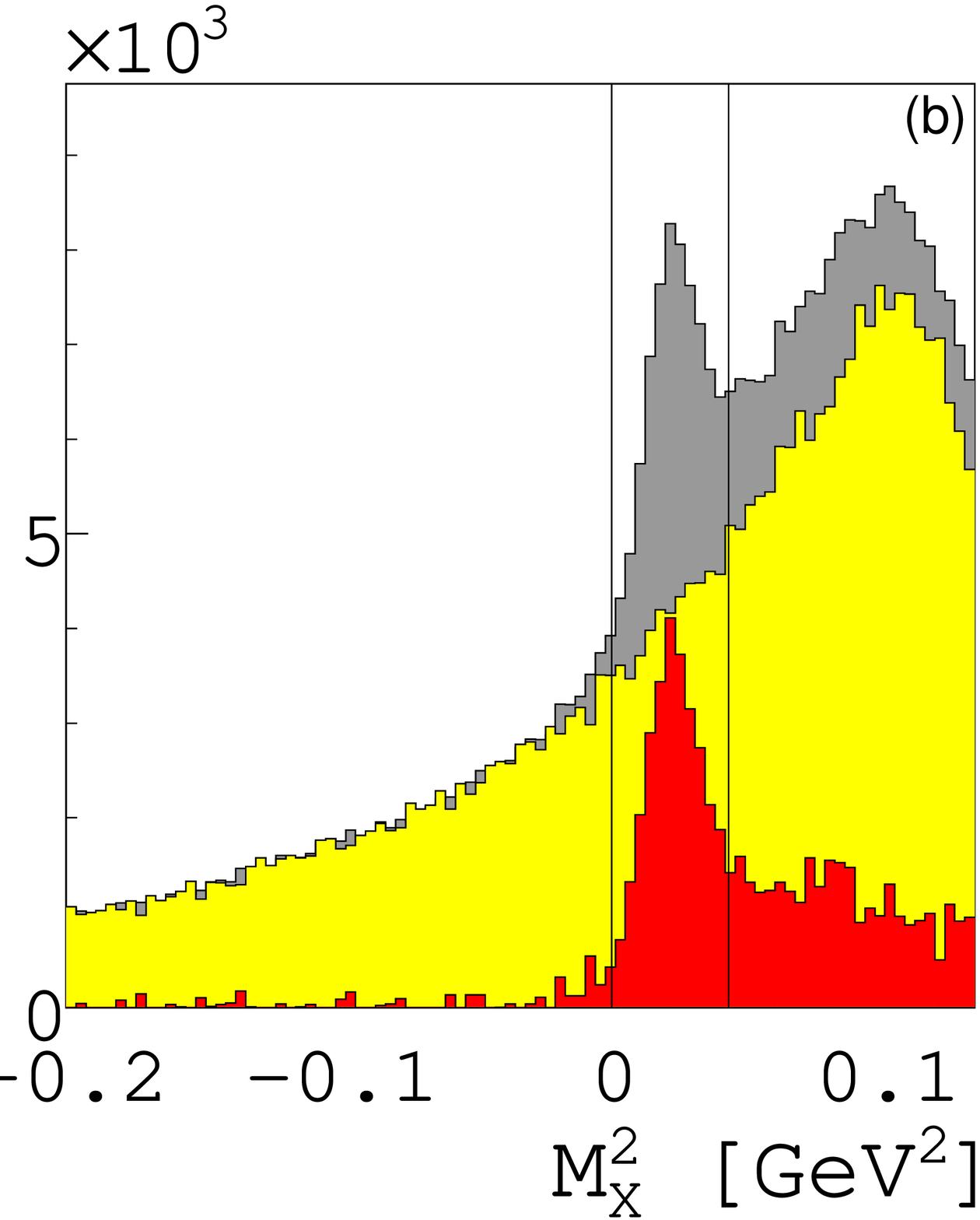}%
 \caption{\label{mm_sample}(Color online) $M_x^2$ spectra of \nh, normalized \carb, and their subtraction for $0<\cos\theta<0.2$ and invariant masses of $1.15\ \gevm <W<1.2\ \gevm$ (a) and $1.4\ \gevm <W<1.5\ \gevm$ (b). The lines show the 2$\sigma$ cut applied to select $\pi^0$ events.}
 \end{figure}

The data contains a large background coming from scattering off  \nitr and the \hel from the target cooling bath. Data on \carb and \hel targets were used to remove these contributions. The procedure to determine the term $C N^{BG}$ consists of two steps. The first step  was to determine how to combine the \carb and \hel data in order to correct for the different amount of \hel displaced by the \nh and \carb targets. Using a calculation based on target thicknesses, densities, and window contributions, the background counts were constructed as $N^{BG}=N^{{\rm {}^{12}C}}-0.22 {\rm N^{{}^4 He}}$. The second step consists of the determination of the normalization constant $C$. The normalization constant is $W$ dependent because of the different ratio of protons to neutrons for the \nitr and \carb targets. The constant $C$ was extracted by normalizing the missing mass spectra of the \nitr and the constructed background to each other in the negative missing mass region, where one has only events stemming from scattering off background nuclei. 
Since the missing mass spectra show a dependence on the center-of-mass polar angle of the proton, $\theta^*$, the normalization constant was extracted for each $W$ and $\cos\theta^*$ bin.
Figure~\ref{mm_sample} shows a sample of missing mass spectra for the \nh, the background, and the subtraction. The figure also shows the cut applied to select neutral pions.

\subsection{Multipion background correction at high $W$}
At higher invariant mass, $W$,  background from the lower missing mass tail of multi-pion events contaminates the $\pi^0$ signal, as can be seen in the right panel of Figure~\ref{mm_sample}. The measured asymmetry, $A_{exp}$, found by combining the number of counts (see Eqs.~\ref{forasymdef} and~\ref{forasymdef_bis}) is therefore given by
\begin{equation}
A_{exp}=f_{\pi^0}A_{\pi^0}+f_{back}A_{back},
\label{eqmultipion}
\end{equation} 
where $A_{\pi^0}$ and $A_{back}$ are the asymmetries of $\pi^0$ and background events, respectively,   $f_{\pi^0}$  is the ratio of $\pi^0$  events to total events, and $f_{back}=1-f_{\pi^0}$ is the fraction of background events.
Equation~\ref{eqmultipion} was used to extract $A_{\pi^0}$. 
$A_{back}$ was measured using the same analysis procedure used for $A_{exp}$, except that events were selected with
a missing mass squared greater than the pion mass squared and less than 0.2 $\gevmsq$.
The values of the background asymmetry $A_{back}$ vary from 1\% to 30\% depending on the kinematic bin. The fraction $f_{back}$ was found by fitting the missing mass spectrum, consisting of the pion peak and the background, with two Gaussian curves and using those to estimate the number of background and pions events in the region selected by the cut.
The fraction was found to be increasing with the invariant mass as expected from a value of 1.2\% for $1.25\ \gevm <W<1.3\ \gevm$ up to 12\% for $1.6\ \gevm <W<1.7\ \gevm$. The overall correction $f_{back}A_{back}$ was from 0.5\% to 2.5\%.

\subsection{$P_eP_t$ determination}
\label{secpept}
\begin{figure}[here,top]
\begin{center}
\includegraphics[width=8.5cm]{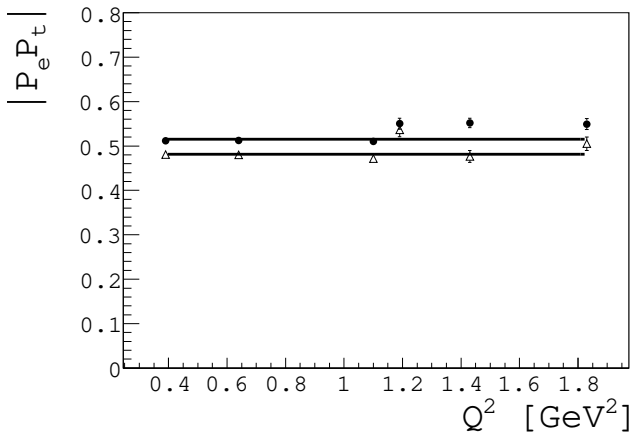}
\end{center}
\caption{The product $|P_eP_t|$ as a function of $Q^2$ for positive (filled circles) and 
negative (open triangles) target polarization runs. The six values for each polarization 
were fitted with a constant in order to obtain the average values 
$P^{}_eP^a_t=0.517 \pm 0.002$ and $P^{}_eP^b_t=- 0.483 \pm 0.003$.}
\label{pept}
\end{figure}

The product of the beam and target polarizations, $P_eP_t$, was determined by comparing the measured elastic asymmetry $A_{meas}$ with the predicted value $A_{theo}$,
\begin{equation}
P_eP_t=\frac{A_{theo}}{A_{meas}},
\end{equation}
where
\begin{equation}
A_{meas}=\frac{\npp-\nmp}{\npp+\nmp}=\frac{P_eP_t\sigma_{et}}{\sigma_0}
\end{equation}
and~\cite{donnelly86}
\begin{widetext}
\begin{equation}
A_{theo}=-\frac{\cos\theta_{\gamma}\sqrt{1-\epsilon^2}+(\frac{Q^2}{4M^2})^{-\frac{1}{2}}\sqrt{2\epsilon(1-\epsilon)}\sin\theta_{\gamma}\cos\phi_{\gamma}\frac{G_E}{G_M}}{\epsilon(\frac{Q^2}{4M^2})^{-1}(\frac{G_E}{G_M})^2+1}.
\end{equation}
\end{widetext}
For the $Q^2$ range in this analysis, $\frac{G_E}{G_M} \sim \mu_p$, where $\mu_p$ is the magnetic moment of the proton ($\mu_p=\frac{1}{2}g$ where $g = 5.5856912$).
The product of beam and target polarization ($P_eP_t$)  was independently determined in six $Q^2$ 
bins and then the average value of these 6 results was calculated.
Figure~\ref{pept} shows the 
results for the positive ($P^{}_eP^a_t$) and negative target polarization data 
($P^{}_eP^b_t$). These measurements allow one to extract target polarizations $P^a_t$, $P^b_t$ 
by simply taking the ratio of these products and the measured beam 
polarization $P_e$.
\subsection{Bin centering corrections}
In order to reduce effects due to the width of
 the kinematic bins in $Q^2$, $W$, $\cos\theta^*$, and $\phi^*$, a bin centering correction was applied to the data. This correction is model dependent and in order to minimize the error, the models that describe the data best were chosen. The Sato and Lee~\cite{lee96,lee01} and MAID07~\cite{drechsel99} models were used for the \delt and the higher $W$ region, respectively.  Neither model includes the data presented in this paper in their fits.
The bin centering correction is defined as:
\begin{equation}
\label{eqbcc}
\beta=\frac{\sigma_{exact}}{\sigma},
\end{equation}
where $\sigma_{exact}$ is the model cross section obtained at the center values of our bins, and $\sigma$ is the model cross section averaged over the bins in the analysis.
This correction was calculated for each beam and target polarization configuration and 
 applied individually to the number of counts $\nppo$, $\nmpo$, $\npmo$, and $\nmmo$ before constructing the asymmetries as described in Eqs.~\ref{forasymdef} and~\ref{forasymdef_bis}.
\subsection{Systematic uncertainties}
Several sources of systematic uncertainties were identified in the analysis procedure. For most cases, the technique used to estimate the size of these uncertainties was to vary individual parameters in the analysis, recalculate the asymmetries, and compare the new results to the original values.

One of the main sources of uncertainties is due to the correction for
nitrogen, which was estimated using a $^{12}$C target.
The statistical uncertainty on the normalization procedure is very small,
however, the whole procedure is based on the assumption that the \carb target spectrum
has a similar shape as  the \nh background.
Based on studies in the elastic region, the normalization factor was found to be known with a precision of 3-4\%. The asymmetry was recalculated using a normalization factor changed by 3.5\%, and the comparison shows a point by point variation of the asymmetry on the order of 4-8\%.

A second source of systematic uncertainties is due to the uncertainties of the beam and target polarization values. The beam polarization is measured with the M{\o}ller polarimeter and it is know with an accuracy of 2\%, which results in a systematic uncertainty on the asymmetry up to 2\%.  The target polarization, extracted from the product $P_eP_t$ as described in section~\ref{secpept}, has a systematic and statistical uncertainty associated with it. The statistical uncertainty 
gives a negligible systematic variation 
of the asymmetry, while the systematic error due to the uncertainty of the ratio $G_E/G_M$ and to the carbon normalization gives a total systematic  uncertainty on the asymmetry of 3.2\%.

Another source of error is the model dependence of the bin centering correction. To study this effect 
a different model, MAID03, was
used for the correction and the difference was assigned as a systematic uncertainty. A point by point variation from 1 to 15\% was found depending on the $W$ range and the asymmetry considered. For the most part these uncertainties were negligible compared to the statistical error.

The multi-pion background correction requires an estimate of the number of events contaminating the pion sample, $f_{back}$. The fraction has both a statistical and systematic uncertainty that contributes to the systematic uncertainty of the asymmetry. The main source of systematic uncertainty was the assumption that the background does not depend on the polar angle of the pion in the center-of-mass, $\theta^*$. An overall uncertainty of 20\% was assigned to $f_{back}$, to account for fluctuations in different $\cos\theta^*$ intervals. A change of this amount causes a point by point variation in the asymmetry up to 10\%, most of the time negligible compared to the statistical uncertainty. 

The effect of cuts was studied as well, in particular the fiducial and missing mass cuts. To study the impact of these cuts on the analysis, the asymmetries were measured using complementary cuts and compared to each other. The two sets of data points were found statistically compatible and therefore no systematic uncertainty was assigned.

The overall systematic uncertainties were found by combining the different contributions in quadrature and are reported point by point in the CLAS Physics Data Base~\cite{clasdb}.
It was found that the systematic uncertainties were smaller than the statistical ones for most of the data points.

\section{Results}
\label{sec:results}
\begin{figure}[here,top]
\begin{center}
\includegraphics[width=8.5cm]{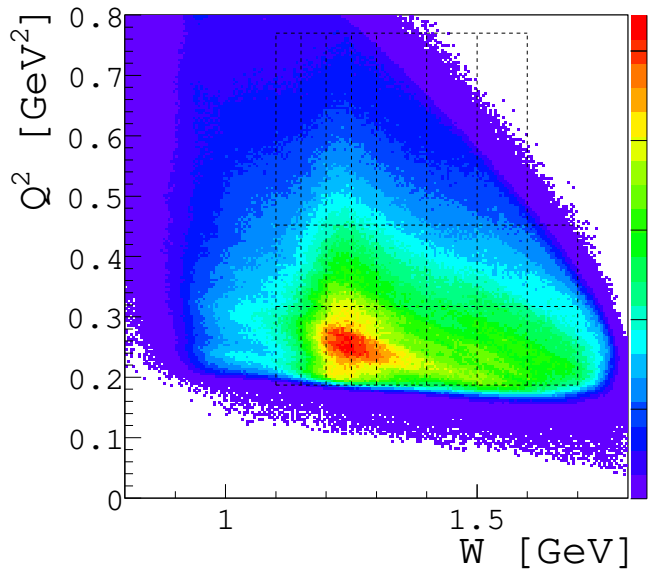}
\end{center}
\caption{\label{kine} (Color online) Kinematic coverage in $Q^2$ and $W$. The dashed lines indicate how the data were subdivided.}
\end{figure}

Target ($A_t$) and double spin ($A_{et}$) asymmetries were measured as a function of the invariant mass $W$, the momentum transfer squared $Q^2$, and the center of mass pion angles, $\theta^*$ and $\phi^*$. The data were divided into 8 bins in $W$, 3 bins in $Q^2$, 10 bins in $\cos\theta^*$, and 15 bins in $\phi^*$. The results consist of 
2,435 data points for each asymmetry, after eliminating bins with nearly zero acceptance.
Table~\ref{Tab:binning} summarizes the binning and Figure~\ref{kine} illustrates the kinematic coverage in $Q^2$ and $W$.
\begin{table}[ht]
\label{tablebin}
\begin{center}
\centering
  \begin{tabular}{|c|c|c|c|}
  \hline
Bin                             & Range             &   Bin size        &  \# bins      \\\hline\hline
                                & 1.1 - 1.3 $\gevm$ &   0.05 $\gevm$    & 4         \\\cline{2-4}
\raisebox{2mm}[2mm][0mm]{$W$}   & 1.3 - 1.7 $\gevm$ &   0.10 $\gevm$    & 4         \\\hline
                                & 0.187 - 0.317 $\gevpsq$   & 0.130 $\gevpsq$   & 1         \\\cline{2-4}
$Q^2$                           & 0.317 - 0.452 $\gevpsq$   & 0.135 $\gevpsq$   & 1         \\\cline{2-4}
                                & 0.452 - 0.770 $\gevpsq$   & 0.318 $\gevpsq$   & 1         \\\hline
$\cos \theta^*$                      & -1.0 - 1.0                & 0.2           & 10        \\\hline
$\phi^*$                        & -180$^{\circ}$ - 180$^{\circ}$                & 24$^{\circ}$      & 15        \\\hline

  \hline
  \end{tabular}
\end{center}
\caption{Binning for the 1.6 $\gev$ data set.}
\label{Tab:binning} 
\end{table}

\begin{figure}
\includegraphics[width=8.5cm]{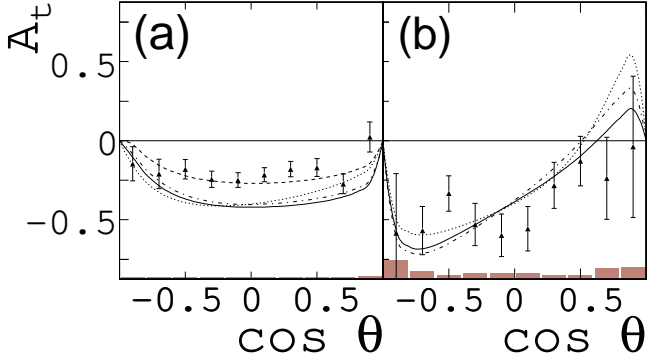}
\caption{\label{Sysatsample} The target asymmetries $A_{t}$ for \qB, $\phi^*$=120$^\circ$, and invariant masses $W=1.225\ \gevm$ (a) and $W=1.35\ \gevm$ (b). Systematic uncertainties are shown as shaded bars.
The curves correspond to the MAID07 (solid), DMT (dotted), Sato and Lee (dashed), and MAID03 (dashed dotted) calculations.}
\end{figure}

\begin{figure}
\includegraphics[width=8.5cm]{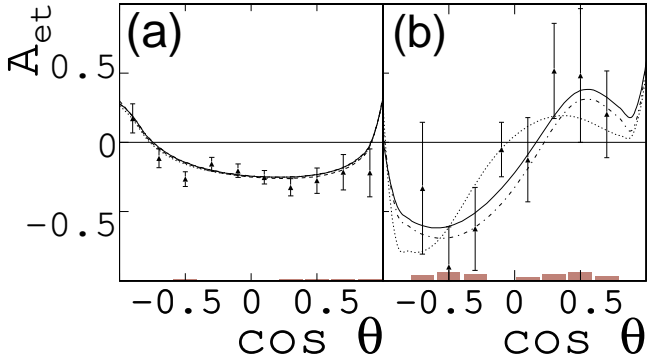}
\caption{\label{Sysaetsample} The target asymmetries $A_{et}$ for \qA, $\phi^*$=120$^\circ$, and invariant masses $W=1.225\ \gevm$ (a) and $W=1.65\ \gevm$ (b). Systematic uncertainties are shown as shaded bars.
The curves correspond to the MAID07 (solid), DMT (dotted), Sato and Lee (dashed), and MAID03 (dashed dotted) calculations. }
\end{figure}

The results were compared with five theoretical approaches: MAID07, MAID03, MAID00, and the DMT model everywhere, and the Sato and Lee model in the \delt region. 
Figures~\ref{Sysatsample} and~\ref{Sysaetsample} show a sample of the results as a function of $\cos\theta^*$. The error bars indicate the statistical uncertainties, while the brown boxes indicate the systematic uncertainties. The complete numerical results are reported in the CLAS Physics Data Base~\cite{clasdb}. We see that the models reproduce well the general trends of the data. The asymmetry signs and the sign changes are generally predicted correctly. Discrepancies are visible at the quantitative level and are different for different models as discussed in Section~\ref{secchisq}. This shows the sensitivity of this data set to the specific model ingredients for the resonant and non-resonant amplitudes in the first, second, and third resonance regions.

\begin{figure}[here,top]
\begin{center}
\includegraphics[width=3.0in]{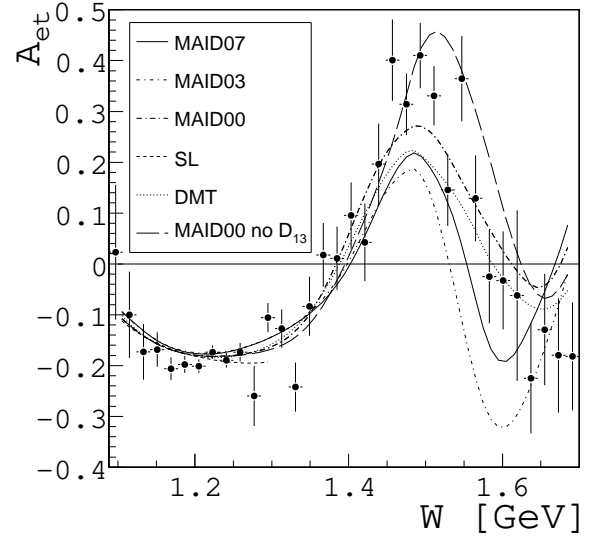}
\end{center}
\caption{\label{Waet} $A_{et}$ as a function of the invariant mass $W$, integrated over the whole range in $\cos\theta^*$, $Q^2$, and $60\deg < \phi^*< 156\deg$. The curves correspond to the MAID07 (solid), DMT (dotted), Sato and Lee (dashed), MAID03 (dashed dotted), MAID00 (long-dashed-dotted) calculations. The long-dashed curve is the MAID00 calculation without the $D_{13}(1520)$ contribution. }
\end{figure}

\begin{figure}[here,top]
\begin{center}
\includegraphics[width=3.0in]{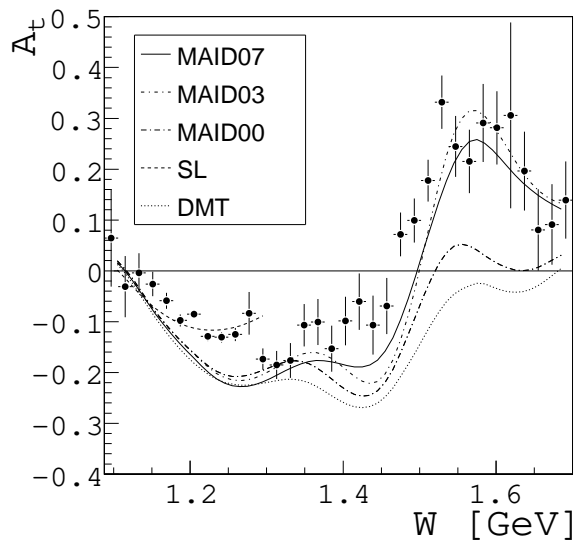}
\end{center}
\caption{\label{Wat} $A_{t}$ as a function of the invariant mass $W$, integrated over the whole range in $\cos\theta^*$, $Q^2$, and $60\deg < \phi^*< 156\deg$. The curves correspond to the MAID07 (solid), DMT (dotted), Sato and Lee (dashed), MAID03 (dashed dotted), and MAID00 (long-dashed-dotted) calculations.}
\end{figure}

\begin{figure*}[here,top]
\begin{center}
\includegraphics[width=4.9in]{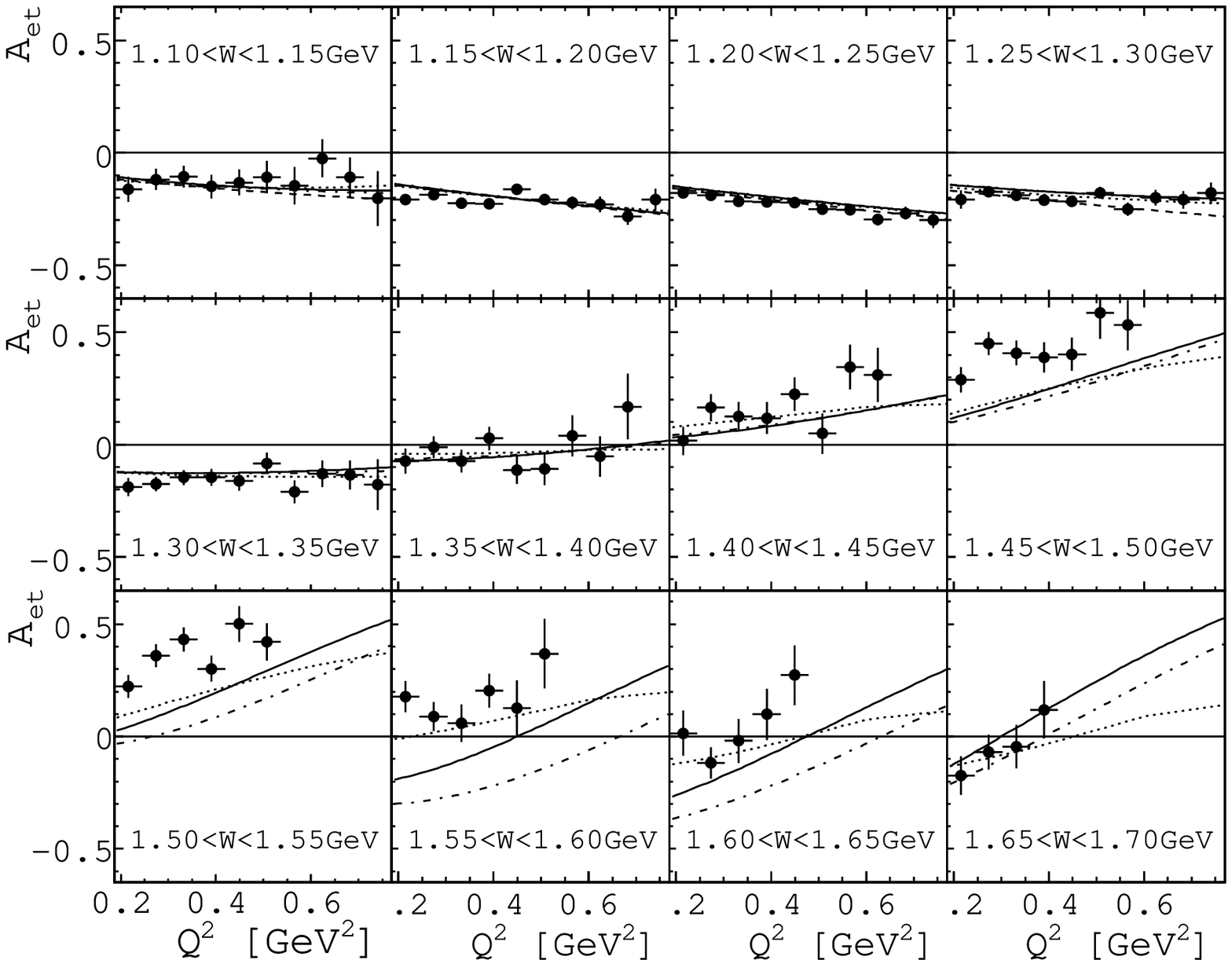}
\end{center}
\caption{\label{Q2aet} $A_{et}$ as a function of the photon virtuality $Q^2$, integrated over the whole range in $\cos\theta^*$, $60\deg < \phi^*< 156\deg$, and 12 invariant mass regions from $1.1\ \gevm$ to $1.7\ \gevm$ in steps of $0.05\ \gevm$.  The curves correspond to the MAID07 (solid), DMT (dotted), Sato and Lee (dashed), and MAID03 (dashed dotted) calculations. }
\end{figure*}

\begin{figure*}[here,top]
\begin{center}
\includegraphics[width=4.9in]{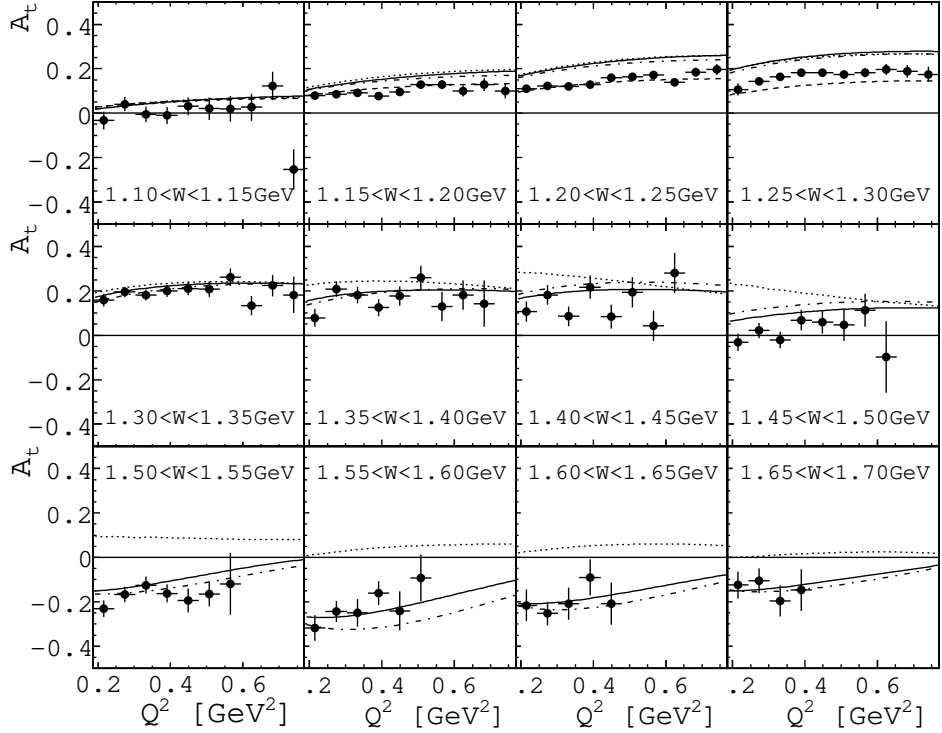}
\end{center}
\caption{\label{Q2at} $A_{t}$ as a function of the photon virtuality $Q^2$, integrated over the whole range in $\cos\theta^*$, $60\deg < \phi^*< 156\deg$, and 12 invariant mass regions from $1.1\ \gevm$ to $1.7\ \gevm$ in steps of $0.05\ \gevm$.  The curves correspond to the MAID07 (solid), DMT (dotted), Sato and Lee (dashed), and MAID03 (dashed dotted) calculations. }
\end{figure*}

Integrated asymmetries were also extracted as a function of the invariant mass $W$ and the momentum transfer $Q^2$. The asymmetries as a function of the invariant mass were calculated integrating over the whole range in $\cos\theta^*$, $Q^2$, and over a partial range in $\phi^*$ from 60$\deg$ to 156$\deg$ to avoid regions with acceptance less than 2\%. Figures~\ref{Waet} and~\ref{Wat} show the results compared to the models. The asymmetries as a function of the momentum transfer were calculated  integrating over the whole range in $\cos\theta^*$, over a partial range in $\phi^*$ from 60$\deg$ to 156$\deg$, and over 
12 bins in $W$ from 1.1 GeV to 1.7 GeV in steps of 0.05 GeV. Figures~\ref{Q2aet} and~\ref{Q2at} show the results compared to the models.

 
Figure~\ref{Q2aet} shows the $Q^2$-dependence 
of the double spin asymmetry $A_{et}$ for 12 bins in $W$. The negative asymmetry and the approximate linear dependence on $Q^2$, as well as the  negative slope observed in the region of the $\Delta(1232)$ resonance are well described by all models.  In the 
mass region of the Roper resonance $P_{11}(1440)$, and the $D_{13}(1520)$ and $S_{11}(1535)$ resonances at $1.45 < W < 1.60$~GeV, the asymmetry is large, positive, and increases 
rapidly with $Q^2$. The DMT and MAID models all underestimate the magnitude. The strength of the $P_{11}(1440)$ in this low $Q^2$ range is known to be small~\cite{Aznauryan2005_1}, and the $S_{11}(1535)$ has been well measured in $\eta$ electroproduction. The excitation strength of the $D_{13}(1520)$ and the relative strength of the $A_{1/2}$ and $A_{3/2}$ amplitudes of this resonance are the main source of uncertainty.  The discrepancy with model parameterizations is likely related to an overestimation of the relative strength of the dominant $A_{3/2}$ amplitude in comparison with the $A_{1/2}$. This sensitivity to the dominant $A_{3/2}$ amplitudes is also shown in Figure~\ref{Waet}, where the magnitude of the computed $A_{et}$ increases without the $D_{13}(1520)$ resonance contribution, bringing it in better agreement with the data in the mass range near 1520 MeV. In the mass range $1.6 < W < 1.7$~GeV the $F_{15}(1680)$ is the dominant resonance. The asymmetry increases rapidly with $Q^2$ and changes sign at $Q^2 \sim 0.35$~GeV$^2$. The MAID07 model approximately describes the $Q^2$dependence, including the sign change. We remark that the observed behavior of $A_{et}$ with $Q^2$ in the region above the $\Delta(1232)$ is consistent with the behavior measured in the $n\pi^+$ channel~\cite{devita01}.

The target asymmetry $A_t$ depends on the imaginary part of interferences involving resonant and non-resonant amplitudes, and its interpretation is less straightforward than the double spin asymmetry. The asymmetry $A_t$ is shown in Figure~\ref{Q2at} as a function of $Q^2$ . There is not a simple explanation of the discrepancies between the models and data for this observable, as the results may depend on the phases of amplitudes that are also strongly energy-dependent. In contrast to the double spin asymmetry, $A_t$ is well described in the mass region of the $\Delta(1232)$ only by the SL model, while the DMT and MAID parameterizations show  significant discrepancies with the data. The higher mass regions are reasonably well described by the MAID parameterizations, while the DMT model shows strong discrepancies, including the wrong sign for the four higher mass bins. 

\subsection{$\chi^2$ comparison}
\label{secchisq}
To evaluate in a more rigorous way how well the models describe the data, a $\chi^2$ comparison was performed.
The $\chi^2$ was defined as:
\begin{equation}
\chi^{2} = \sum_{i} \frac{(x_{i}^{\rm data} - x_{i}^{\rm model})^{2}}{(\sigma_{i}^{\rm data})^{2}},
\end{equation}
where $x_{i}^{\rm data}$ is the value of each experimental point for all the asymmetries and $x_{i}^{\rm model}$ is the corresponding value of the theoretical prediction. Since the model is given without errors, only the experimental statistical uncertainties $\sigma_{i}^{\rm data}$ were used in the denominator.
The comparison gave the results listed in Table~\ref{chisq}.
\begin{table}[here,top]
\begin{center}
\begin{tabular}{ c| c| c| c| c}
{\sl Model}     & $A_{t}$   & $A_{et}$ & $A_{t}$ & $A_{et}$  \\ \hline
& \multicolumn{2}{c|}{\tiny $W < 1.3\ \gevm$} & \multicolumn{2}{c}{\tiny $W > 1.3\ \gevm$}\\
& \multicolumn{2}{c|}{\tiny $\text{ndf} = 1306$ } & \multicolumn{2}{c}{\tiny $\text{ndf} = 1129$}\\ \hline
{MAID07}        &    1.98         & 1.04  & 0.99 & 1.30\\
{MAID03}        &    1.75         & 1.05  & 1.20 & 1.86\\
{MAID00}        &    1.97         & 1.03  & 1.97 & 1.07\\
{SL}                 &    1.00         & 1.05  &  -   & - \\
{DMT}             &    2.09         & 1.02  & 1.71 & 1.17\\\end{tabular}
\end{center}
\caption{\boldmath $\chi^2$ per number of degree of freedom (ndf) comparison between the data and the five models.}
\label{chisq}	
\end{table}
The calculation led us to make a few conclusions about the models. First of all,
in the \delt region, all the models are in good agreement with the measured double  spin asymmetry, $A_{et}$. This is not surprising since $A_{et}$ is dominated by
the resonance multipoles (i.e. $|M_{1+}|^2$, $\reopmop$)~\cite{drechsel92} that dominate
the unpolarized cross section. Since the model fits are based on unpolarized cross 
sections, this agreement is expected. At higher $W$ none of the models give a consistently good description of the
data for both the target and the double spin asymmetry. The MAID07 
parameterization shows improvements for $A_t$ over previous versions, while $A_{et}$ is
not as well reproduced as by the older MAID00 version. At this level it is not
possible to draw further conclusions and only including these data in a new global fit for
these models will give a better understanding of the sensitivity of the data and
possibly identify the origin of the discrepancies.

\section{Summary}
\label{sec:summary}
In this paper we have presented a set of longitudinal target polarization asymmetries $A_t$ and beam-target double
polarization asymmetries $A_{et}$  for the reaction $ep\to ep\pi^0$ covering the first, second, and part of the third nucleon
resonance regions in a wide range of azimuthal and polar center-of-mass angles for the final state $\pi^0$. The data also
cover a range in photon virtuality $Q^2$.  For the first time, target polarization asymmetries have been measured in the
resonance region and in the mass range above the $\Delta(1232)$ resonance. The new data complement the already
published results obtained in the $\Delta(1232)$ region~\cite{biselli03} for different kinematics. Longitudinal target
asymmetries and beam-target asymmetries were measured in 2,435 bins each. The polarization asymmetries show
strong sensitivity to both resonant and non-resonant amplitudes and are highly selective of  model parameterizations for
amplitudes underlying the $\pi^0$ electroproduction process. Due to the large number of data points, only samples of specific
kinematics and integrated quantities are compared with model predictions. 
Although the models agree well with the double spin asymmetries in the $\Delta(1232)$ mass region, three of the four models surveyed have a substantial disagreement beyond the $\Delta(1232)$ region and all models show a significant disagreement with the single spin target asymmetry data.  
The discrepancies, when quantified using the statistical interpretation from Ref.~\cite{pdg} of the $\chi^2$ values in Table~\ref{chisq} above, amount to less than a 5\% probability that the measured double spin asymmetry data set ($A_{et}$) would be consistent with any of the given models for $W$ above the $\Delta(1232)$.  For the single spin target asymmetry ($A_t$) the probability is less than 59\%. 
The full impact of the complete data set on
our understanding of the nucleon structure in the regime of strong QCD and confinement can only be obtained by including
these new data sets in global analyses that incorporate all exclusive pion electroproduction cross sections and polarization
observables. This effort is currently underway, and is the subject of forthcoming publications~\cite{aznauryan08}.
The full set of data is available at the CLAS Physics Data Base~\cite{clasdb}.       

\section{Acknowledgments}

We would like to thank the staff of the Accelerator 
and Physics Divisions at the Jefferson Laboratory for 
their outstanding efforts to provide us with the high 
quality beam and the facilities for the data analysis. 
This work was supported by the U.S. Department of Energy and the National Science Foundation, the French 
Commissariat a lÕEnergie Atomique, the Italian Istituto 
Nazionale di Fisica Nucleare, and the Korean Science and 
Engineering Foundation. The Southeastern Universities 
Research Association (SURA) operates the Thomas Jefferson National Accelerator Facility for the United States 
Department of Energy under Contract No. DE-AC05-84ER40150.

\end{document}